\documentclass[namedreferences,hyperref,optionalrh]{spr-sola}
\usepackage{graphicx}        
\usepackage{color}           

\usepackage{url}
\usepackage{multirow}
\usepackage{tabularray}
\usepackage{booktabs}
\usepackage{longtable}
\usepackage{array}
\usepackage[table]{xcolor}
\usepackage{hyperref}
\hypersetup{
    colorlinks,
    linkcolor={red!50!red},
    citecolor={red!50!red},
    urlcolor={red!80!red}
    }

\chardef\us=`\_

\begin{document}

\begin{frontmatter}
\title{On the Penumbra-to-Umbra Ratio from 1660 to 1676}

\author[addressref=aff1,corref,email={ned@geo.phys.spbu.ru}]{\inits{N.V.}\fnm{Nadezhda}~\snm{Zolotova}\orcid{0000-0002-0019-2415}
}\author[addressref={aff2},email={mikhail.vokhmianin@oulu.fi}]{\inits{M.V.}\fnm{Mikhail}~\lnm{Vokhmyanin}\orcid{https://orcid.org/0000-0002-4017-6233}}

\address[id=aff1]{St. Petersburg State University, Universitetskaya nab. 7/9, 198504 St. Petersburg, Russia}

\address[id=aff2]{Space Climate Group, Space Physics and Astronomy Research Unit, University of Oulu, Oulu, Finland}

\runningauthor{N.V.~Zolotova, M.V.~Vokhmyanin}
\runningtitle{On the Penumbra-to-Umbra Ratio}

\begin{abstract}

Sunspot engravings made in the Maunder minimum are used to evaluate the fine structure of sunspots. Based on 78 images of the full solar disk and 77 images of individual sunspots, we have evaluated the ratio of penumbral-to-umbral area (P/U) to be $4.2 \pm 2.5$ and $3.8 \pm 2.3$, respectively. These results are consistent with previous estimates before, during, and after the Maunder minimum, as well as with the P/U ratio observed in the largest sunspot groups in solar cycle 24. This suggests that the near-surface convection mode has most likely remained unchanged since the early seventeenth century. We also found that schematic sunspot drawings tend to underestimate the P/U ratio.

\end{abstract}
\end{frontmatter}

\section{Introduction}
     \label{S-Introduction} 

The ratio of penumbral and umbral areas is one of the statistical quantities of sunspots and has been studied since \citet{1933PASP...45...51N} reported the P/U ratio of 5.7. \citet{1939MiZur..14..439W} studied stable sunspots and found that the ratio of the diameters of penumbra and umbra varies considerably decreasing with the penumbra size with the most common value of $\approx 2.4$. Since then, various measures have been utilized to evaluate sunspot structure: the ratio of penumbral-to-umbral radius or diameter, the umbra/penumbra ratio, and the ratio of the total sunspot area to umbral area. A substantial survey of these studies can be found in \citet{2003A&ARv..11..153S, 2013SoPh..286..347H, 2024SoPh..299...19C}.

Various dependencies between the P/U ratio and such sunspot parameters as latitudinal distribution, size, magnetic field strength, and also the phase of the solar cycle have been studied \citep{1971BAICz..22..352A, 1991BAICz..42..316A, 2006Ap&SS.306...23J, 2014SoPh..289.1013B, 2019SoPh..294...72J, 2022RAA....22i5012H, 2023SoPh..298...63T}. Sunspot group data from the Royal Greenwich Observatory, Debrecen Solar Observatory, and Kodaikanal Solar Observatory are widely utilized in this regard.

The P/U ratio in the regular sunspot groups is on average between 5 and 6 \citep{2013SoPh..286..347H, 2018SoPh..293..104C}. \citet{2022FrASS...919751J} confirmed this result for individual sunspots and reported that for the regular sunspots the P/U ratio has been a fairly constant value over the last century in Kodaikanal photographic data. According to observations from the Kislovodsk Mountain astronomical station in cycle 24, \citet{2019SoPh..294...45T} found that the relative area of the umbra significantly depends on the type of sunspot: pore, transitional sunspot, or largest area sunspot in a group.

\citet{1976IAUS...71....3P} hypothesized that convection was weakened during the Maunder minimum, resulting in a lower pressure on the penumbra. Therefore, \citet{1997rscc.book.....H} expected higher P/U values throughout the global solar minimum. \citet{1993SoPh..146...49B} reported that the global contrast of sunspots, which in turn affects the total and spectral solar irradiance, significantly depends on the P/U ratio. Taken as a whole, a long chain of solar-terrestrial relationships results in a correlation between the penumbra-to-umbra ratio and surface temperature anomalies \citep{1979ClCh....2...79H}.

In this paper, we estimate the P/U ratio from the observational data during the period of 1660\,--\,1676. We utilized the sunspot parameters recently reconstructed by \citep{2025SoPh..300...17Z}. All processed sunspot images are available at \url{http://geo.phys.spbu.ru/~ned/History.html}.

\section{Data}
\label{S-Data}

\begin{figure}    
\centerline{\includegraphics[width=1\textwidth,clip=]{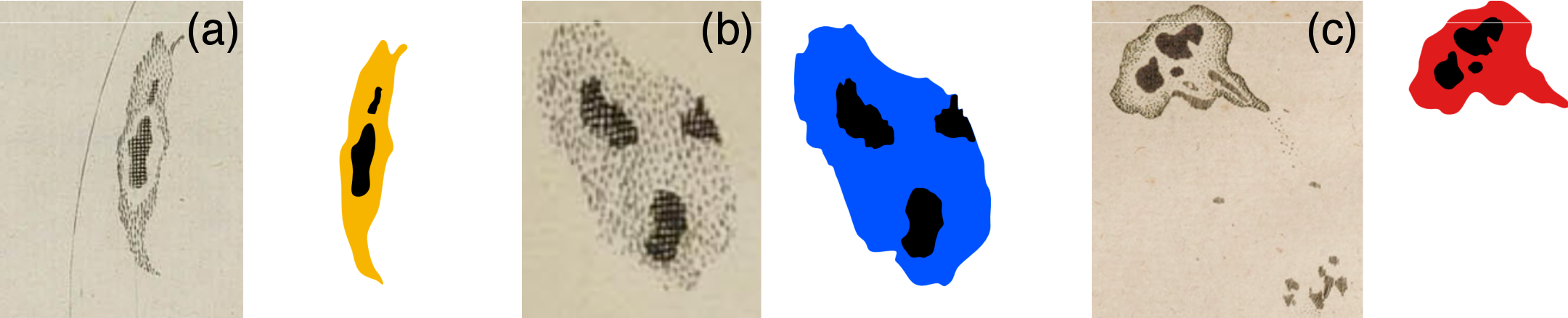}}
\small
        \caption{Examples of sunspot groups on the original engravings and their sunspot reproductions: (\textbf{a}) the near-limb sunspot from \citet{Cassini_1671_2}; (\textbf{b}) sunspot from \citet{Cassini_1672}; (\textbf{c}) sunspot group from \citet{Monnier_1741}.}
\label{Fig1}
\end{figure}

In this work, we determine the ratio of the penumbral area to umbral area for individual sunspots. We define an individual sunspot as consisting of one or more umbrae enclosed by a common penumbra. Sunspot engravings usually present the umbra and penumbra as a dashed and dotted texture making it difficult to estimate their areas. Figure~\ref{Fig1} shows the original images and their reproductions created using common image-editing software. By setting a color intensity threshold, we measure the umbral and penumbral areas either in pixels for individual engravings of sunspots or in millions of the visible solar hemisphere (msh) for the sunspots on the solar disk. \textbf{In the latter case, each pixel is weighted based on its angular distance.}

\begin{figure}    
\centerline{\includegraphics[width=1\textwidth,clip=]{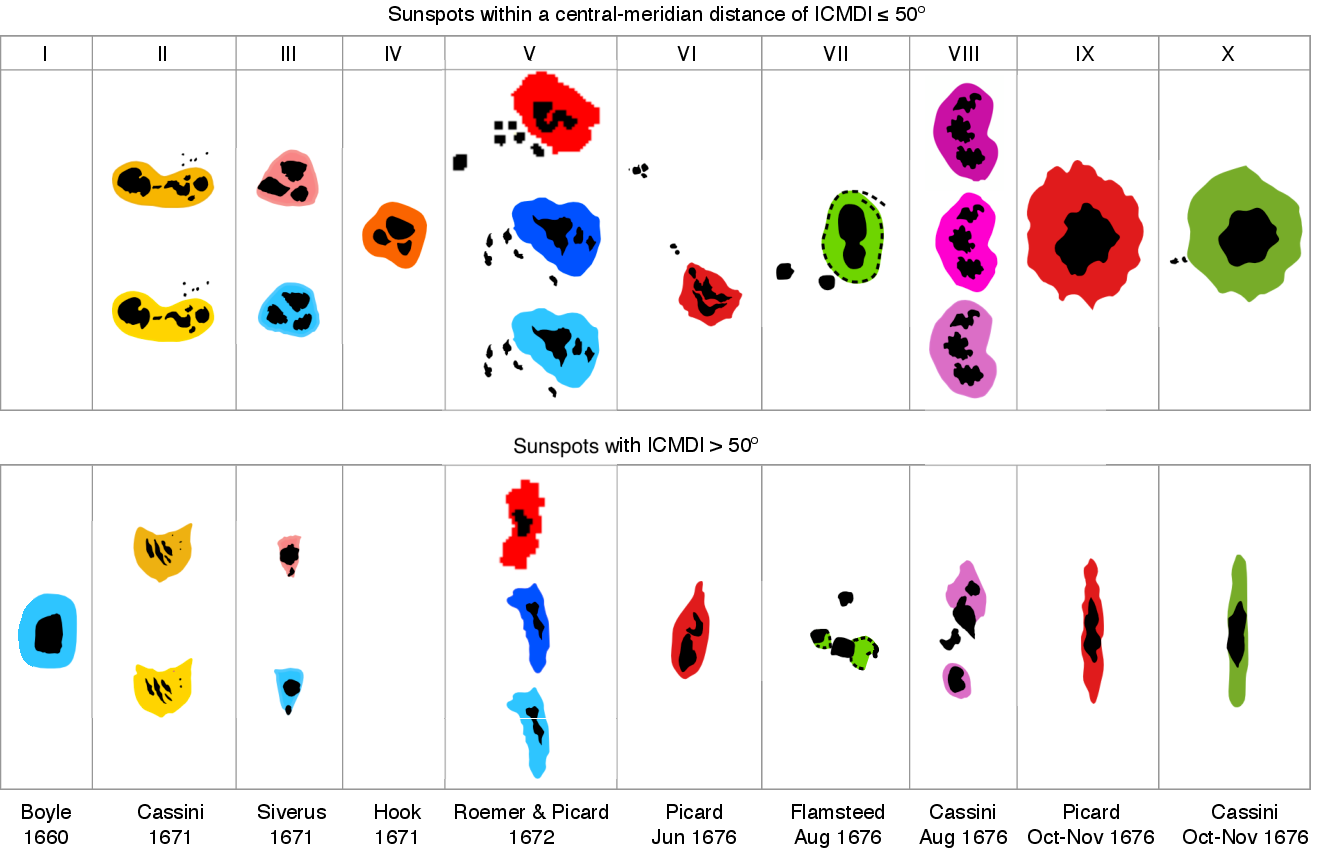}}
\small
        \caption{Examples of sunspot groups located away from the limb (central meridian distance is less than $50^{\circ}$, top row) and near the limb (central meridian distance greater than $50^{\circ}$, bottom row), whose individual sunspots have been analyzed in this study.}
\label{Fig2}
\end{figure}

Figure~\ref{Fig2} shows several examples of sunspot groups, whose individual sunspots are analyzed in this work. Sunspots located far from the limb are presented in the top row, and the bottom row shows their appearance when they were near the limb. The latter are found to be less reliable. We typically use a central-meridian distance (CMD) threshold of $50-60^{\circ}$. Depending on the series of observations, the drawing/engraving may already be less accurate at $|CMD| = 49^{\circ}$, or may still be reliable at $|CMD| = 62^{\circ}$. In such cases, we subjectively decide on the relevance of the image.

The first example in Figure~\ref{Fig2} is a schematic sunspot made on 7 May 1660 by Robert Boyle and published by \citet{1671RSPT....6.2216C}.

The second column in Figure~\ref{Fig2} represents samples of a sunspot group observed by Jean-Dominoque Cassini and published in two French monographs: one covering 11\,--\,13 August \citep{Cassini_1671}, and the other spanning 14\,--\,20 August 1671 \citep{Cassini_1671_2}. These samples are illustrated in dark-yellow. The sunspot group from the monographs reprinted in the Philosophical Transactions \citep{Cassini_1671_Ph_Tr, Cassini_1671_Ph_Tr_2} are painted in light-yellow. 

The third column in Figure~\ref{Fig2} represents schematic sunspot engravings observed by Heinrich Siverus from 18 August to 15 September 1671, but reported by Martin \citet{Fogelius_1671_Ph_Tr, Fogelius_1671}: sunspots with the pink-colored penumbra are from the letter to Henry Oldenburg, and the blue-colored ones were published in the Philosophical Transactions. The fourth column shows the same sunspot according to the only schematic drawing on 11 September 1671 by \citet{Hook_1671}. 

The fifth column contains the sunspot group from the observations initiated by Ole Christensen Roemer (Rømer) and continued in collaboration with Jean Picard from 18 October to 22 November 1672. The sunspots with the red-colored penumbra are from the solar disk engraving, and the dark-blue ones are the individual engravings of the sunspot group, both from \citet{Cassini_1672}. The sunspots with the light-blue penumbra belong to the individual engravings of the sunspot group from \citet[reprinted from 1699]{Bion_1751}. 

The sixth column shows the individual engravings of a sunspot group presumably tracked by Picard from 26 June to 4 July 1676, but published by \citet{Monnier_1741}.

The next column shows sunspots from the schematic solar disk drawing by John Flamsteed from 6 to 14 August 1676 \citep{Flamsteed_1676_Ph_Tr}. In the original engraving, the sunspot penumbra is shown by the dashed curve. Note the discontinuity of the penumbra curve which is the case for some sunspots. Flamsteed mentioned this, confirming that this is not an engraving inaccuracy. However, this peculiarity introduces an uncertainty in measuring the penumbra area. The green area shows the region by which we calculate the penumbra area for these two cases. 

The eighth column is the schematic sunspots made on 8 and 14 August 1676 by \citet{Cassini_1676_Ph_Tr, Cassini_1676}. All three copies of these drawings that we found show the outer boundary of the penumbra indistinctly, which adds ambiguity in further calculations. 

The ninth column represents samples of the individual engravings of the sunspot group presumably observed by Picard from 30 October to 30 November 1676 \citep{Monnier_1741}. The last column contains the individual engravings of the same sunspot group from \citet{Cassini_1676_Nov}. For comparison, Picard's and Cassini's engravings for the same days are shown: 24 November at the top, and 30 November 1676 at the bottom. Picard's engraving skips two pores (small sunspots without penumbra). \citet{Cassini_1671_Ph_Tr, Cassini_1671_Ph_Tr_2} classified tiny objects beyond the penumbra boundary as black points rather than sunspots. Note that all the observers mentioned above defined a sunspot by its umbra, while referring to a penumbra as a musty or darkish cloud, corona, or nubecula (a cloud in Latin). More details on the original data can be found in \citet{2025SoPh..300...17Z}.

If an observer recorded measurements of the solar diameter, linear size of a sunspot group, and a distance from the solar limb, it is possible to project an individual engraving of the sunspot onto the solar disk with an accuracy up to the orientation of the sunspot group. We performed the sunspot mapping for the observations made by Boyle, Cassini, Hook, Picard and Roemer, and Picard.

\begin{figure}    
\centerline{\includegraphics[width=0.5\textwidth,clip=]{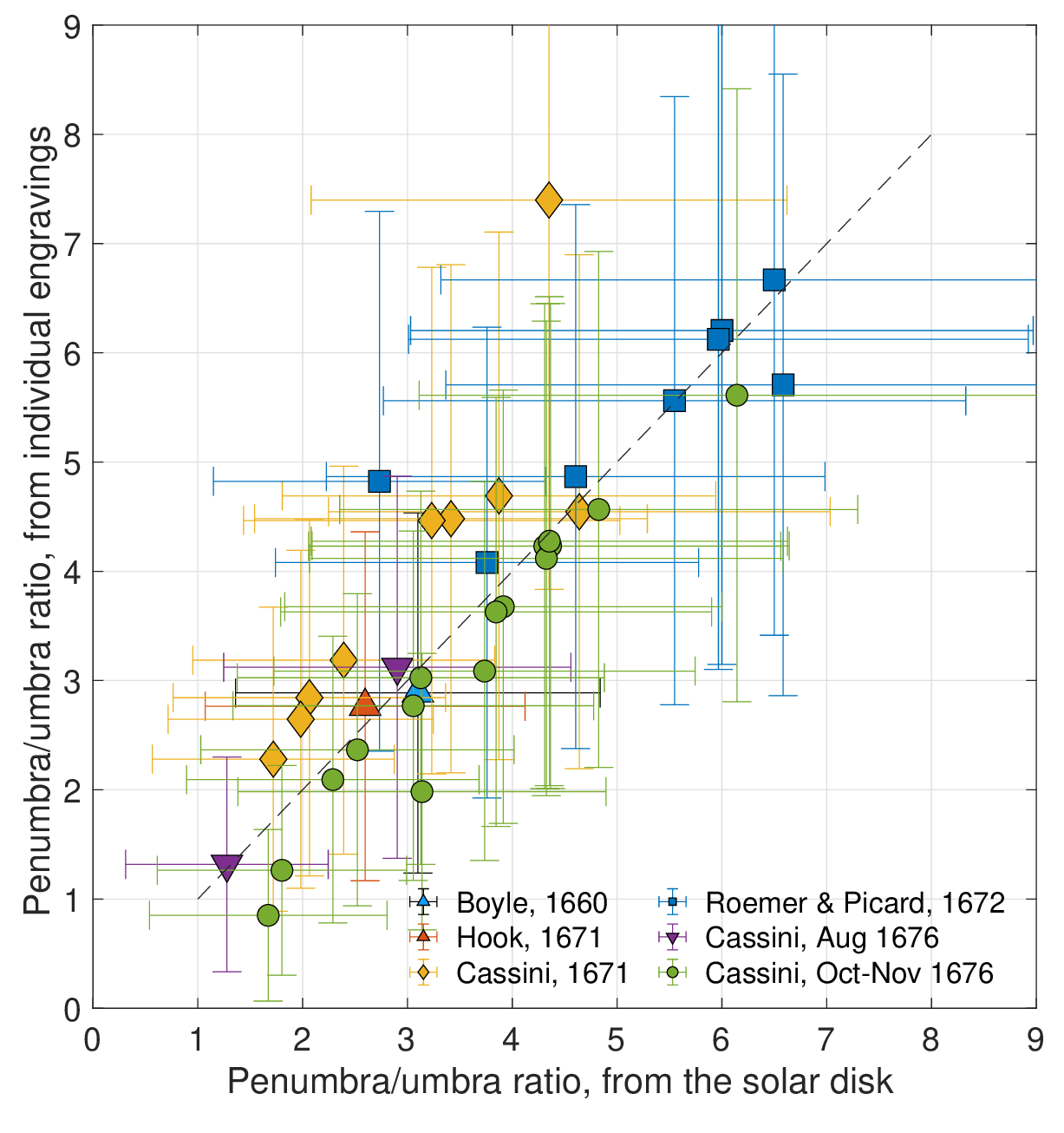}}
\small
        \caption{Relationship between the penumbra-to-umbra area ratio determined from individual drawings and engravings of sunspots and the ratio determined after we mapped these sunspots onto the solar disk.}
\label{Fig3}
\end{figure}

Ideally, the penumbra-to-umbra ratio determined from the individual drawings and engravings of sunspots and the ratio determined for these sunspots mapped onto the solar disk (see \citet{2025SoPh..300...17Z} for details) must be identical. Figure~\ref{Fig3} illustrates this relationship. We found the best consistency of the calculated P/U ratios in the observations made by Boyle, Hook, and Cassini in August 1676. In Cassini's observations in 1671, the penumbra often occupies a larger portion of the sunspot in the individual engravings compared to the full solar disk images. In contrast, in Cassini's observations in October\,--\,November 1676 the P/U ratios are sometimes lower for the sunspots on the solar disk drawings. The largest divergence is exhibited by the two sunspots, which are affected by foreshortening and sunspot size. Due to the projection effect, each pixel of the solar disk image has a different area (in msh) which depends on its angular distance. Closer to the limb the effect of foreshortening becomes stronger, thus variations in number of pixels covered by umbra and penumbra result in larger differences of P/U. Another factor affecting the P/U ratio is the sunspot size. For small sunspots with the umbra size less then 10~msh, the P/U becomes less reliable as umbra size can vary significantly depending on the intensity threshold. We therefore calculated the P/U for both individual sunspot engravings and sunspots mapped onto the solar disc.

\section{Results}
\label{S-Results}

\begin{figure}    
\centerline{\includegraphics[width=1\textwidth,clip=]{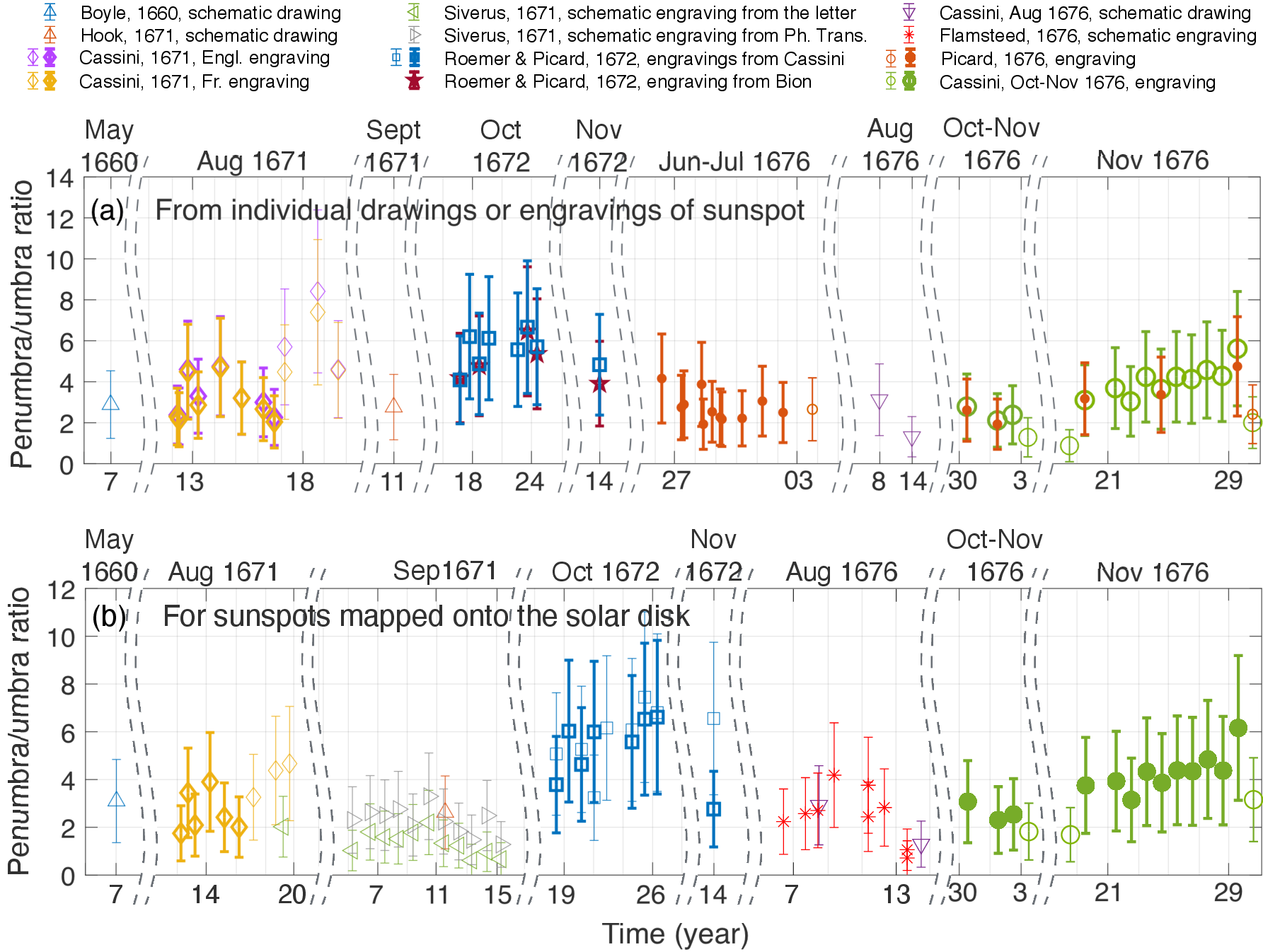}}
\small
        \caption{Evolution of the penumbra-to-umbra area ratio for all sunspots analyzed in this work.\textbf{ The x-axis represents the time of observations. Large time gaps are indicated with dashed curves.} (\textbf{a}) the ratio determined from individual drawings or engravings of sunspots, (\textbf{b}) the ratio determined for sunspots mapped onto the solar disk. \textit{Bold symbols} indicate the more reliable values derived from the detailed engraving of sunspot away from the limb. \textit{Light symbols} indicate unreliable values from schematic representation of sunspot or sunspot near the solar limb.}
\label{Fig4}
\end{figure}

Figure~\ref{Fig4} represents the evolution of the penumbra-to-umbra area ratio for all sunspots analyzed in this work. These values are divided into two types: the first one consists of more reliable values derived from the detailed images of a sunspot away from the limb (bold symbols); the second one is made up of less reliable values obtained from the schematic images of a sunspot or sunspot images near the solar limb (light symbols). Figure~\ref{Fig4}a illustrates the P/U ratio examined from the individual drawings and engravings of sunspots, and Figure~\ref{Fig4}b illustrates the ratio for sunspots mapped onto the solar disk. Note that the foreshortening effect does not have a consistent impact on the portion of penumbra in the sunspot area; it can sometimes increase it (the P/U rises in Cassini's observations in 1671) or decrease it (the P/U falls in Cassini's observations in 1676).

\begin{table}
\caption{Statistics of the P/U ratio using the Monte-Carlo simulations}
\label{Tab1}
\begin{tabular}{c|ccc}     
\toprule          
\multicolumn{2}{r}{}& \textbf{Individual} & \textbf{Solar} \\
\multicolumn{2}{r}{} & \textbf{engravings} & \textbf{disk} \\
\hline 
\multicolumn{2}{r}{}  & \textbf{Mean $\pm$ std. dev.} & \textbf{Mean $\pm$ std. dev.} \\
\multicolumn{2}{c}{\textbf{Observer}} & \textbf{Median} & \textbf{Median} \\
\multicolumn{2}{c}{\textbf{(number of observations)}}  & \textbf{95\% CI} & \textbf{95\% CI} \\
\toprule
& &  3.4 $\pm$ 1.9 & \\
& Cassini, 1671, Engl. Engr. &  3.10 & \\
 &  (8 obs.)&  2.24\,--\,4.75 & \\
 \cline{2-4}
 & & 3.5 $\pm$ 2.0 & 2.7 $\pm$ 1.6 \\
 &Cassini, 1671, Fr. Engr.&  3.10 & 2.50 \\
 &  (8 obs. \& 6 obs.)&  2.28\,--\,4.69 & 1.72\,--\,3.87 \\
\cline{2-4}
\multirow{8}{*}{{\rotatebox[origin=c]{90}{\textbf{Reliable sunspot images}}}}  & & 5.1 $\pm$ 2.6  &  \\
&Roemer \& Picard, 1672, by Bion& 5.08 & \\
&  (5 obs.)& 3.9\,--\,6.46 & \\
\cline{2-4}
 & & 5.7 $\pm$ 2.8  & 5.4 $\pm$ 2.9 \\
 &Roemer \& Picard, 1672, by Cassini& 5.51 & 5.03 \\
 &  (8 obs. \& 8 obs.)& 4.08\,--\,6.67 & 2.73\,--\,6.58 \\
\cline{2-4}
 & & 3.0 $\pm$ 1.7  &  \\
 & Picard, 1676& 2.82 & \\
 &  (11 obs.)& 1.91\,--\,4.74 & \\
\cline{2-4}
 & & 3.8 $\pm$ 2.1 & 3.5 $\pm$ 1.9 \\
 & Cassini, Oct\,--\,Nov 1676& 3.58 & 3.32 \\
 &  (13 obs. \& 13 obs.)& 2.09\,--\,5.61 & 2.29\,--\,4.31 \\
\cline{2-4}
 & \cellcolor[HTML]{ebf1fc}&  \cellcolor[HTML]{ebf1fc}\textbf{3.8 $\pm$ 2.3} & \cellcolor[HTML]{ebf1fc} \textbf{4.2 $\pm$ 2.5} \\
 & \cellcolor[HTML]{ebf1fc} \textbf{Full dataset}& \cellcolor[HTML]{ebf1fc} \textbf{3.46} & \cellcolor[HTML]{ebf1fc}\textbf{3.76} \\
 & \cellcolor[HTML]{ebf1fc}& \cellcolor[HTML]{ebf1fc} \textbf{1.91\,--\,6.47} &  \cellcolor[HTML]{ebf1fc} \textbf{1.77\,--\,6.57} \\
\toprule   
& Boyle, 1660  (1 obs.)& **2.9 $\pm$ 1.7 & **3.0 $\pm$ 1.7 \\
 \cline{2-4}
& Hook, 1671 (1 obs.) & **2.8 $\pm$ 1.6 & **2.6 $\pm$ 1.5 \\
\cline{2-4}
& & & 1.6 $\pm$ 1.0 \\
& Siverus, 1671, Lett.& & 1.41\\
& (12 obs.)& & 0.59\,--\,2.17 \\
\cline{2-4}
\multirow{8}{*} {{\rotatebox[origin=c]{90}{\textbf{Unreliable sunspot images}}}}  & & & *2.4 $\pm$ 1.4 \\
& Siverus, 1671, Ph. Trans.& & 2.25\\
&  (11 obs.)& & 1.25\,--\,3.27 \\
\cline{2-4}
& & & 6.0 $\pm$ 3.0 \\
& Roemer \& Picard, 1672, by Cassini& & 5.81\\
&  (9 obs.)& & 3.21\,--\,7.42 \\
\cline{2-4}
& & & *2.7 $\pm$ 1.7 \\
& Flamsteed, 1676& & 2.43\\
&  (9 obs.)& & 0.7\,--\,4.17 \\
\cline{2-4}
& & **2.3 $\pm$ 1.5 & **2.1 $\pm$ 1.3\\
& Cassini, Aug 1676& -- & --\\
& (2 obs. \& 2 obs.)& -- &  --\\
\cline{2-4}
&\cellcolor[HTML]{fcece4} & \cellcolor[HTML]{fcece4} *\textbf{2.7 $\pm$ 1.6} & \cellcolor[HTML]{fcece4} *\textbf{3.1 $\pm$ 2.5} \\
& \cellcolor[HTML]{fcece4} \textbf{Full dataset}& \cellcolor[HTML]{fcece4} \textbf{2.47} & \cellcolor[HTML]{fcece4}\textbf{2.40}\\
&\cellcolor[HTML]{fcece4} & \cellcolor[HTML]{fcece4} \textbf{1.25\,--\,3.03} & \cellcolor[HTML]{fcece4} \textbf{0.62\,--\,7.05} \\
 \bottomrule
\end{tabular}\\
$*$ \footnotesize{Probability density function for this dataset has not a normal distribution;}\\
$**$ \footnotesize{Due to one or two data-points, the Monte-Carlo simulations were not applied.}
\end{table}

Since the number of high-quality sunspot drawings that include penumbra depictions is very limited, we used the Monte-Carlo simulation approach to estimate the P/U ratios based on different sunspots in the analyzed period. For each sunspot observation in the series, we generate additional points from the normal distributions, where the mean and sigma values are equal to the P/U ratio and its error estimated for each sunspot separately. Table~\ref{Tab1} provides the mean, standard deviation, median, and a 95\% confidence interval (95\% CI) of the P/U ratio in each series of observations found using all simulated observations. Overall, the P/U ratio is estimated to be $3.8 \pm 2.3$ for the individual engravings and $4.2 \pm 2.5$ for the solar disk, \textit{i.e.}, the area of the penumbra is about 80\% of the entire sunspot. The full dataset of subjectively lower quality images yields the average P/U ratio of $2.7 \pm 1.6$ and $3.1 \pm 2.5$, respectively, \textit{i.e.}, the area of the penumbra is 72\,--\,75\% of the entire sunspot.

\begin{figure}    
\centerline{\includegraphics[width=1\textwidth,clip=]{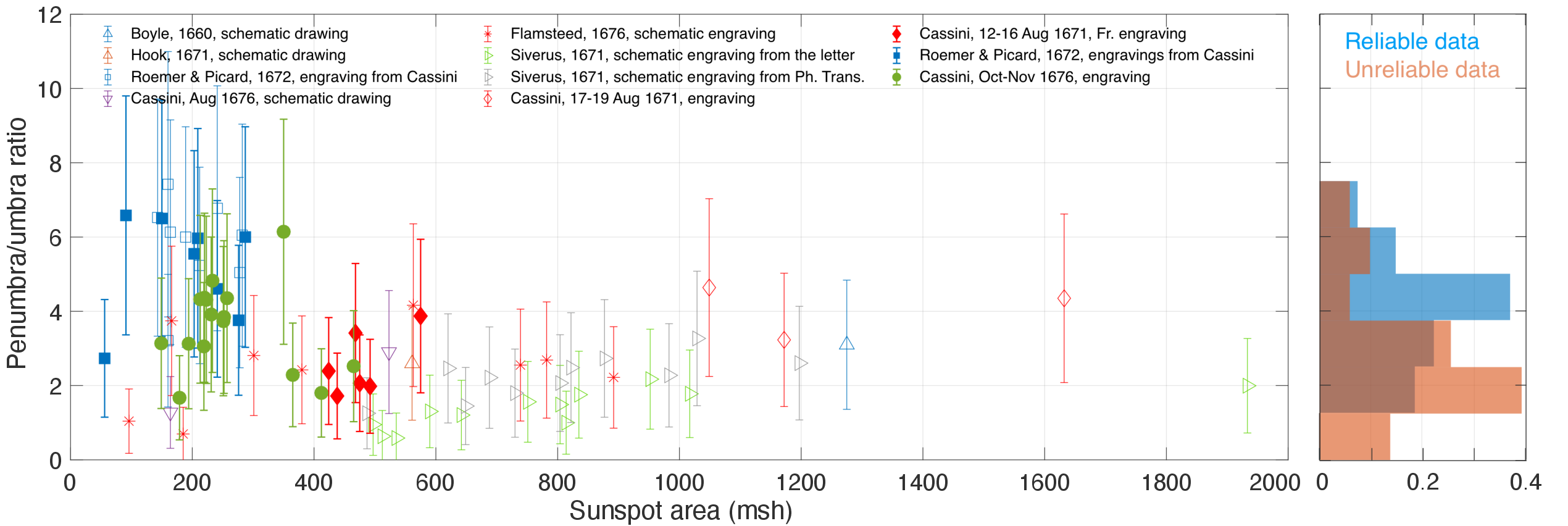}}
\small
        \caption{Penumbra-to-umbra ratio as a function of total sunspot area. \textit{Filled symbols} indicate detailed representation of sunspot away from the limb defined as reliable data. \textit{Unfilled symbols} indicate schematic representation of sunspot or sunspot near the solar limb defined as unreliable data. (\textit{On the right}) Probability density function of reliable and unreliable data.}
\label{Fig5}
\end{figure}

Figure~\ref{Fig5} represents the P/U ratio as a function of the total sunspot area (umbra plus penumbra). Reliable values indicated with filled symbols notably correspond to smaller sunspots, while less reliable estimates are based on the schematic images (unfilled symbols) which tend to overestimate total sunspot areas. Yet, all these sunspots appeared to be regular, i.e., having an area of several hundred msh.

We assume that the sunspots, analyzed in this study, are those with the largest area within their sunspot group. The remaining sunspots (except Flamsteed's, Figure~\ref{Fig2}) can be classified as pores because they were drawn as small ($<15$ msh) sunspots without penumbra. Note that these trailing sunspots cumulatively affect the penumbra-to-umbra ratio calculated for the sunspot group in the observation by Roemer and Picard, Picard in June 1676, and Flamsteed: it is 30\,--\,50\% less than the ratio obtained for individual sunspots. On contrary, these ratios are equal in the observations by Boyle, Hook, Siverus, and Picard and Cassini, both in October\,--\,November 1676.

Earlier, \citet{2018ApJ...865...88C} provided the U/P ratio based on observations during the Maunder minimum, which we can compare with the values obtained from the reliable subset of Cassini's observations in 1676. For the sunspot from 30 October to 3 November 1676, \citeauthor{2018ApJ...865...88C} obtained $1.8 \pm 0.7$ (inverted U/P value) as compared to our estimates of $2.1 \pm 1.3$ for the actual observation ($2.3 \pm 1.3$ for the Monte-Carlo simulated data). For the sunspot observed from 19 to 29 November 1676, \citeauthor{2018ApJ...865...88C} obtained $3.2 \pm 1.0$, while we got $3.9 \pm 2.1$ ($4.0 \pm 2.2$). The Monte-Carlo test yields slightly higher values because the probability density function of the scarce P/U ratio measurements is not perfectly normally distributed. Commonly, the above results are compatible within the errors. Notice that for comparison, we only subtracted the individual engravings of sunspots. The lack of clear boundaries between the umbra and penumbra (Figure~\ref{Fig1}) also has an effect. A 5\%-uncertainty in the areas of the penumbra and umbra would result in a 10\% bias in the P/U ratio. Furthermore, the difference between the engravings of different printing houses affects the ratio (Figure~\ref{Fig4}).

\section{Discussion}
\label{S-Discussion}

Since the Maunder minimum became a representative grand solar minimum, various theoretical models have proposed simulations to reproduce the disappearance of visible (surface) activity. These models include solar-like dynamo with nonlinearities in the velocity field, $\Lambda$-quenching, irregularities and pumping within the inductive action arising from turbulent convection, $\alpha$-quenching, nonlinear parity modulation, interplay of the different convection modes, and \textit{etc.} \citep[and references therein]{1999A&A...343..977K, 2010ApJ...724.1021K, 2010LRSP....7....3C, 2016A&A...589A..56K}. Despite these efforts, the exact trigger of the grand minima remains unknown.

Tracking sunspot rotation can shed light on the global velocity field during the Maunder minimum \citep{1993A&A...276..549R}, and the P/U ratio can indicate notable changes, if any, in the convection regime near the surface. Theoretically, the weak-convection mode \citep{1976IAUS...71....3P} initiates a quenching of the turbulent pressure on sunspots, which entails an increase in the penumbra size. Therefore, \citet{1997rscc.book.....H} expected higher P/U values throughout the global solar minimum.

For sunspots before the Maunder minimum, in 1612, \citet{2021SoPh..296....4V} estimated the average P/U to be $3.9 \pm 0.7$ for the detailed engravings made by Galilei. Throughout the Maunder minimum, \citet{2018ApJ...865...88C} estimated the P/U ratio to be $3.7 \pm 1.1$. These values are consistent with our new results for the subset of reliable data, $3.8 \pm 2.3$ from individual engravings and $4.2 \pm 2.5$ for the entire solar disk drawings. For the reliable subset of regular sunspots drawn by Johann Christoph Müller soon after the Maunder minimum in 1719\,--\,1720, \citet{2022Ge&Ae..62..845Z} yield the average P/U of 4.0 (CI 95\%: 1.9\,--\,8.5).

Thus, despite the relatively poor statistics, the P/U ratio, based on reliable sunspot images, remains fairly persistent before, throughout, and just after the Maunder minimum, ranging from 3.8 to 4.2. This estimate is fairly close to that obtained for the largest sunspots within sunspot group during solar cycle 24 \citep{2019SoPh..294...45T}, which was 4.3, and shown to be independent of the cycle phase. Around the maximum of cycle 21, \citet{1990SoPh..129..191B} estimated the P/U to be 4.17 for small individual sunspots and 3.13 for the large ones. This suggests that the near-surface convection mode has most likely remained unchanged since the early seventeenth century.

We found that schematic sunspot drawings tend to lower the P/U ratio: $2.1 \pm 0.7$ for Cigoli's drawings in 1612, $3.1 \pm 2.0$ for Eustachio Manfredi and $3.4 \pm 1.7$ for Maria Eimmart in 1703, 3.7 (CI 95\%: 1.9\,--\,8.5) in 1719\,--\,1720, and $2.7 \pm 1.6$ from individual engravings and $3.1 \pm 2.5$ from the solar disk drawings, as obtained in this study.

The P/U ratio obtained from historical sunspot reports suffers from uncertainties such as the discrepancy between the engravings of different printing houses or enlarged pores accompanying the largest area sunspot. In general, the P/U ratio calculated for sunspot groups is therefore less reliable than that calculated for individual sunspots.

\section{Conclusions}
\label{S-Conclusions}

In this work, we analyzed sunspot drawings and engravings made in the Maunder minimum over the period of 1660\,--\,1676, by Boyle, Cassini, Siverus, Hook, Flamsteed, Picard, and Roemer and Picard. Overall, we analyzed 77 individual images of the sunspots and 78 drawings of the entire solar disk. Concerning the last one, part of them were taken from the original manuscripts, and part of them were reconstructed with accuracy up to the orientation of a sunspot group.

The individual sunspots considered here for the evaluation of the penumbral-to-umbral ratio are regular and apparently the largest within the group. The results were obtained separately for subjectively reliable P/U values, derived from the detailed images, and for unreliable P/U values, the one from schematic images and sunspots observed near the limb. 

To evaluate basic statistical quantities, the Monte-Carlo simulations were employed. For the group of reliable estimates, the average P/U ratio is $3.8 \pm 2.3$ for individual engravings and $4.2 \pm 2.5$ for the solar disk, \textit{i.e.}, the area of the penumbra is about 80\% of the entire sunspot. These results are consistent with \citet{2019SoPh..294...45T} who studied the largest-area sunspot in a group in cycle 24. Our findings agree within the error with the ratios obtained around the Maunder minimum \citep{2018ApJ...865...88C,2022Ge&Ae..62..845Z}. The persistence of the penumbra-to-umbra ratio before, during, and just after the Maunder minimum, as well as in the modern epoch, suggests that there has been no noticeable change in the surface convection regime.

We found that schematic drawings underestimate the P/U ratio as sunspots accompanying the largest-area sunspot may have been drawn at an enlarged scale or misidentified by an observer as something other than a sunspot \citep{Cassini_1671}. Including these drawings in the overall statistics could thus alter the P/U value by dozens of percent.

\begin{acks}

We used data from the archives of Royal Society Publishing \url{https://royalsocietypublishing.org/journals}, Bibliotheque Nationale de France \url{https://gallica.bnf.fr}, Biodiversity Heritage Library \url{https://www.biodiversitylibrary.org}, and Bibliotheque l’Observatoire de Paris \url{https://bibnum.obspm.fr}, and from our sunspot catalog \url{https://geo.phys.spbu.ru/~ned/History.html}.

\end{acks}

\section*{Funding}
This work has no funding.

\section*{Disclosure of Potential Conflicts of Interest}
The authors declare that they have no conflicts of interest.

\bibliographystyle{spr-mp-sola}
\bibliography{Zolotova_Vokhmyanin}  

\IfFileExists{\jobname.bbl}{} {\typeout{}
\typeout{****************************************************}
\typeout{****************************************************}
\typeout{** Please run "bibtex \jobname" to obtain} \typeout{**
the bibliography and then re-run LaTeX} \typeout{** twice to fix
the references !}
\typeout{****************************************************}
\typeout{****************************************************}
\typeout{}}

\end{document}